\documentclass[twoside,12pt]{article}
\usepackage{epsfig}

\def\Journal#1#2#3#4{{#1} {#2} (#4) #3 }

\def\PRL{\em Phys. Rev. Lett.}

\def\PREP{\em Phys. Rep.}

\def\PRD{{\em Phys. Rev.} D}

\newcommand{\be}{\begin{equation}}
\newcommand{\ee}{\end{equation}}
\newcommand{\bea}{\begin{eqnarray}}
\newcommand{\eea}{\end{eqnarray}}

\begin{document}

\title{ \vspace{1cm} Neutrino Flavor Ratio on Earth and at Astrophysical Sources}
\author{Kwang-Chang Lai,$^{1,2}$ Guey-Lin Lin,$^{1,2}$ and T.C Liu$^{1,2}$\\
\\
$^1$Institute of Physics, National Chiao-Tung University, Hsinchu 300, Taiwan\\
$^2$Leung Center for Cosmology and Particle Astrophysics,\\
National Taiwan University, Taipei 106, Taiwan.}
\maketitle
\begin{abstract}
 We present the reconstruction of neutrino flavor ratios at astrophysical sources.
 For distinguishing the pion source and the muon-damped source to the 3$\sigma$ level,
 the neutrino
 flux ratios, $R\equiv\phi(\nu_\mu)/(\phi(\nu_e)+\phi(\nu_\tau))$
 and $S\equiv\phi(\nu_e)/\phi(\nu_\tau)$, need to be measured in
 accuracies better than 10$\%$.

\end{abstract}
\section{Introduction}
Astrophysical neutrino sources are characterized by their neutrino
flavor ratios. For example, the pion source generates the neutrino
flavor ratio,
$\{\phi(\nu_e):\phi(\nu_\mu):\phi(\nu_\tau)\}=\{1:2:0\}$, where
neutrinos are produced by pion decays and the subsequent decays of
muons. In the muon-damped source, the neutrino flux produced by muon
decays are suppressed due to interactions of muons with strong field
or matter \cite{Rachen,Kashti,Kachelriess}, and the flavor ratio of
muon-damped source is
$\{\phi(\nu_e):\phi(\nu_\mu):\phi(\nu_\tau)\}=\{0:1:0\}$.
 Due to the neutrino oscillation effect, the
flavor ratio observed on Earth is different from that at the
astrophysical source. It is possible that two rather different
sources may generate almost the same flavor ratio on Earth after
neutrino oscillations. Neutrino telescope measures the flux-ratio
parameters $R\equiv\phi(\nu_\mu)/(\phi(\nu_e)+\phi(\nu_\tau))$ and
$S\equiv\phi(\nu_e)/\phi(\nu_\tau)$. In this talk, we discuss the
discrimination of astrophysical neutrino sources, taking into
account the uncertainties on neutrino mixing angles, $CP$ violation
phase and achievable accuracies for determining $R$ and $S$.


\section{Statistical Analysis}
The initial neutrino flux and the observed neutrino flux on the
Earth is connected by the probability matrix $P$ via
\begin{eqnarray}
\left(
  \begin{array}{c}
     \phi(\nu_e) \\
    \phi (\nu_{\mu}) \\
    \phi (\nu_{\tau})\\
  \end{array}
\right)
 =
\left(
   \begin{array}{ccc}
     P_{ee} & P_{e\mu} & P_{e\tau} \\
     P_{\mu e} & P_{\mu\mu} & P_{\mu\tau} \\
     P_{\tau e} & P_{\tau\mu} & P_{\tau\tau} \\
   \end{array}
 \right)
 \left(
  \begin{array}{c}
     \phi_0(\nu_e) \\
    \phi_0(\nu_{\mu}) \\
    \phi_0(\nu_{\tau})\\
  \end{array}
\right)\equiv P\left(
  \begin{array}{c}
     \phi_0(\nu_e) \\
    \phi_0(\nu_{\mu}) \\
    \phi_0(\nu_{\tau})\\
  \end{array}
\right)
, \label{source_earth}
\end{eqnarray}
The matrix elements $P_{ij}$ are functions of neutrino mixing angles
and $CP$ violation phase. Since the distance between the source and
earth is sufficiently large, the matrix element $P_{ij}$ does not
depend on the neutrino mass-squared differences nor depend on the
neutrino energy. This study adopts the following global fitting
result of neutrino mixing angles \cite{Gonzalez}
\begin{eqnarray}
\sin^2\theta_{12}=0.32^{+0.02}_{-0.02}, \,
\sin^2\theta_{23}=0.45^{+0.09}_{-0.06}, \, \sin^2\theta_{13}< 0.019.
\label{bestfit}
\end{eqnarray}

\begin{figure}
\includegraphics[scale=0.23]{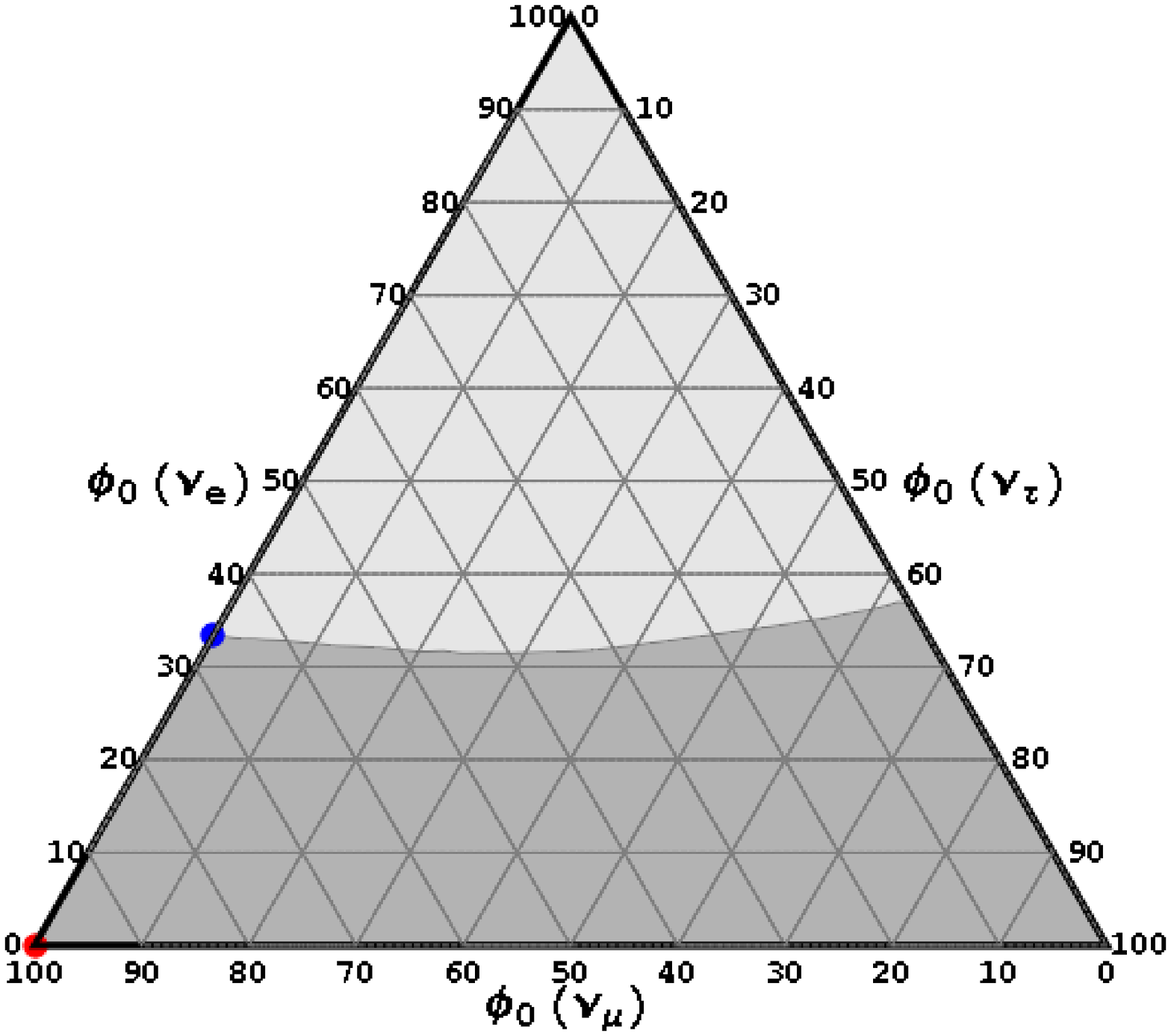}
\includegraphics[scale=0.23]{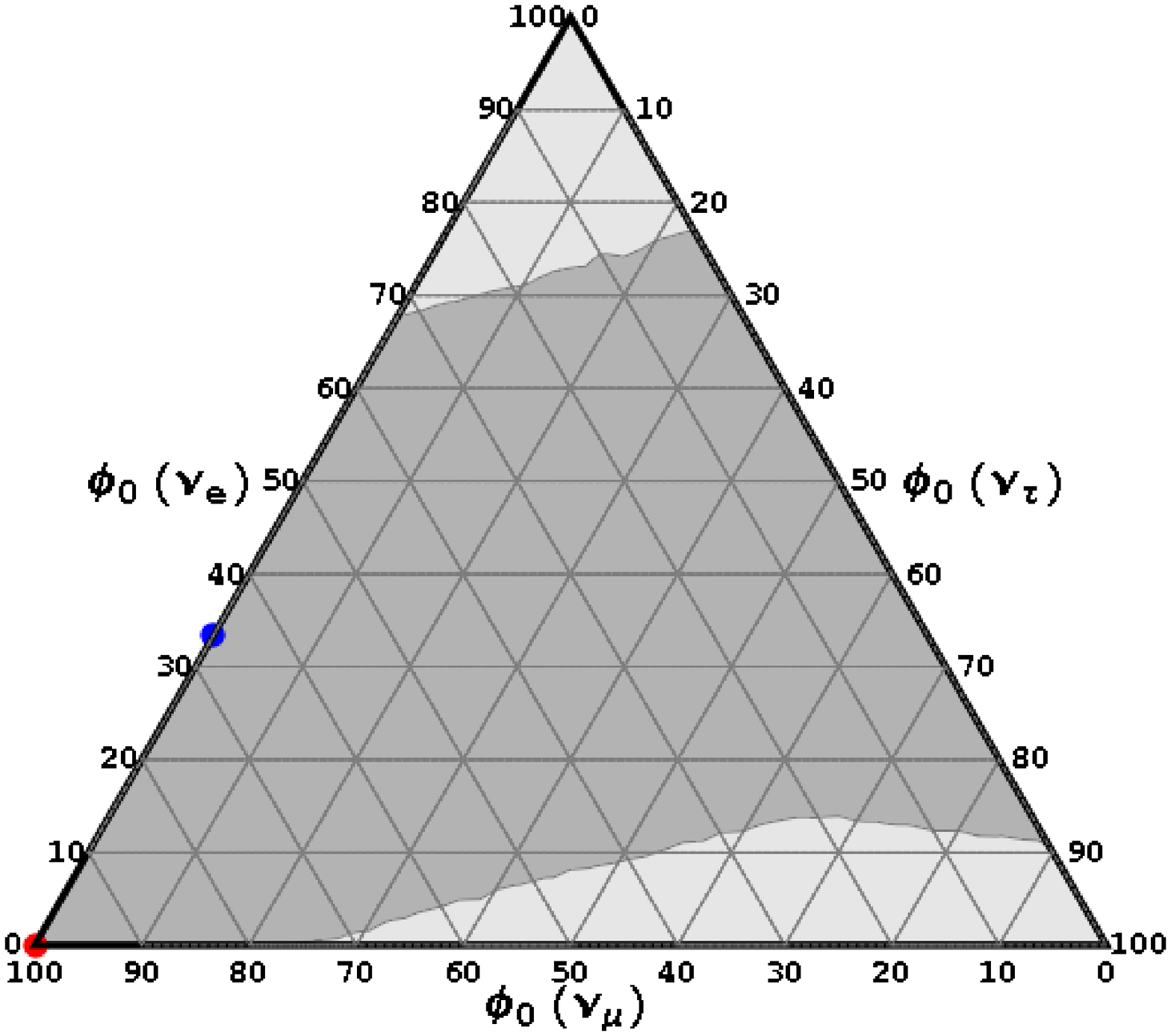}
\includegraphics[scale=0.23]{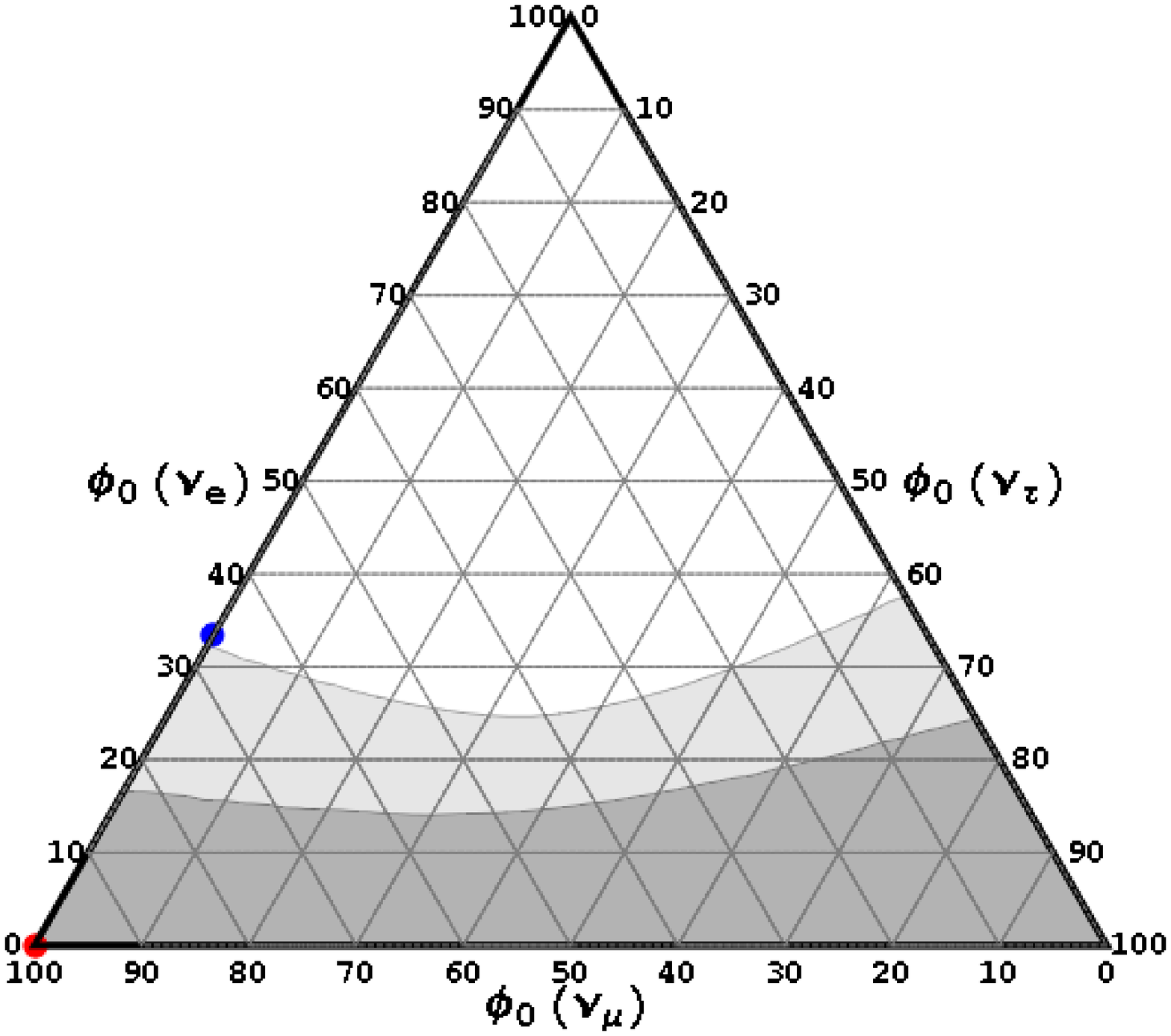}
\includegraphics[scale=0.23]{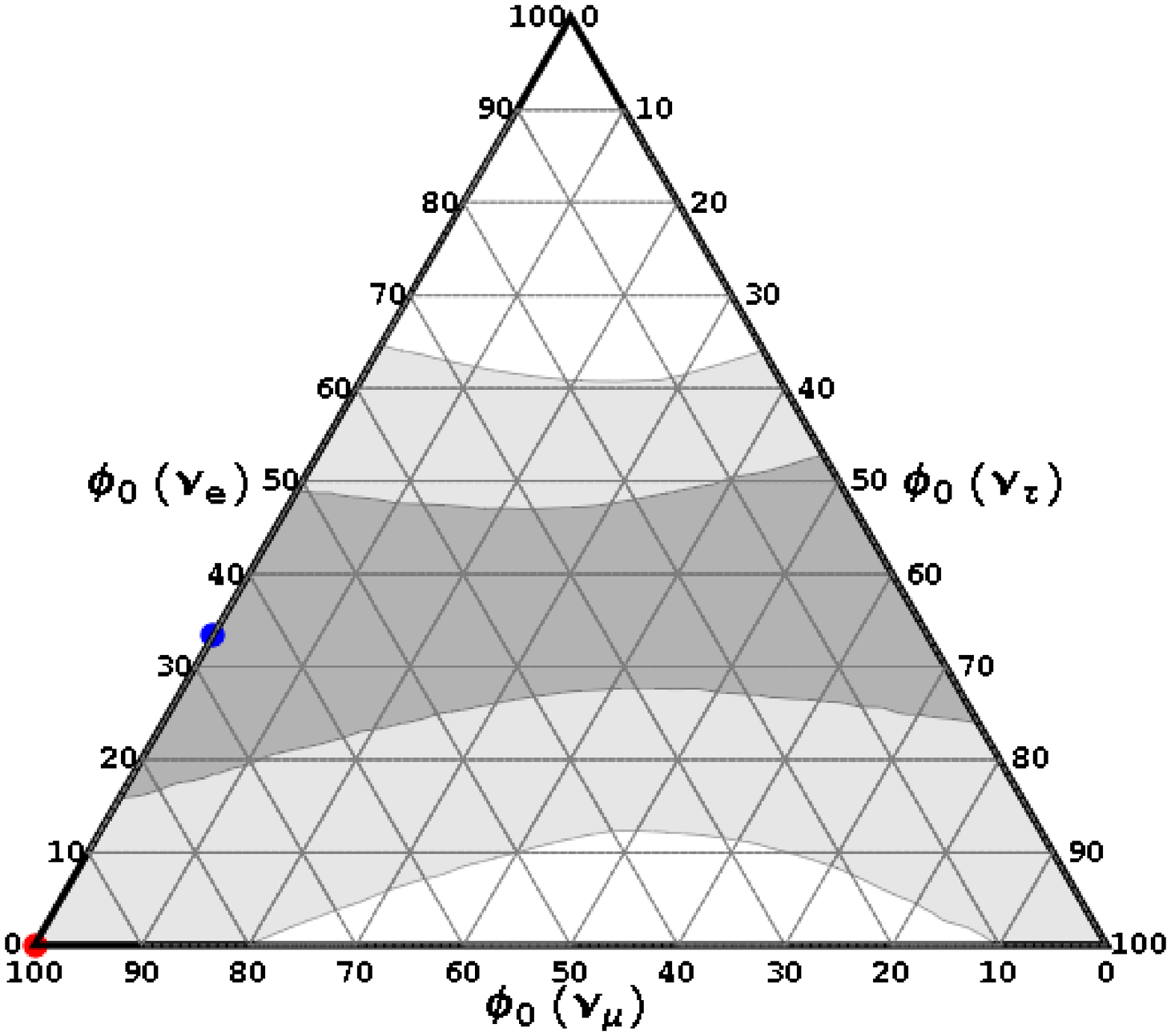}
\caption{The reconstructed ranges of neutrino flavor ratios. The red
point marks the muon-damped source
$\{\phi(\nu_e):\phi(\nu_\mu):\phi(\nu_\tau)\}_\mu=\{0:1:0\}$, the
blue one marks the pion source
$\{\phi(\nu_e):\phi(\nu_\mu):\phi(\nu_\tau)\}_p=\{1/3:2/3:0\}$. Gray
and light gray areas denote the reconstructed $1\sigma$ and
$3\sigma$ ranges. The first two panels are results for an input
muon-damped source and an input pion source, respectively, with only
$R$ measured ($\Delta R/R=10\%$). The next two panels are the
results for an input muon-damped source and an input pion source,
respectively, with $\Delta R/R=10\%$ and $\Delta S/S=12\%$.}
\label{fig1}
\end{figure}

\begin{figure}
\includegraphics[scale=0.23]{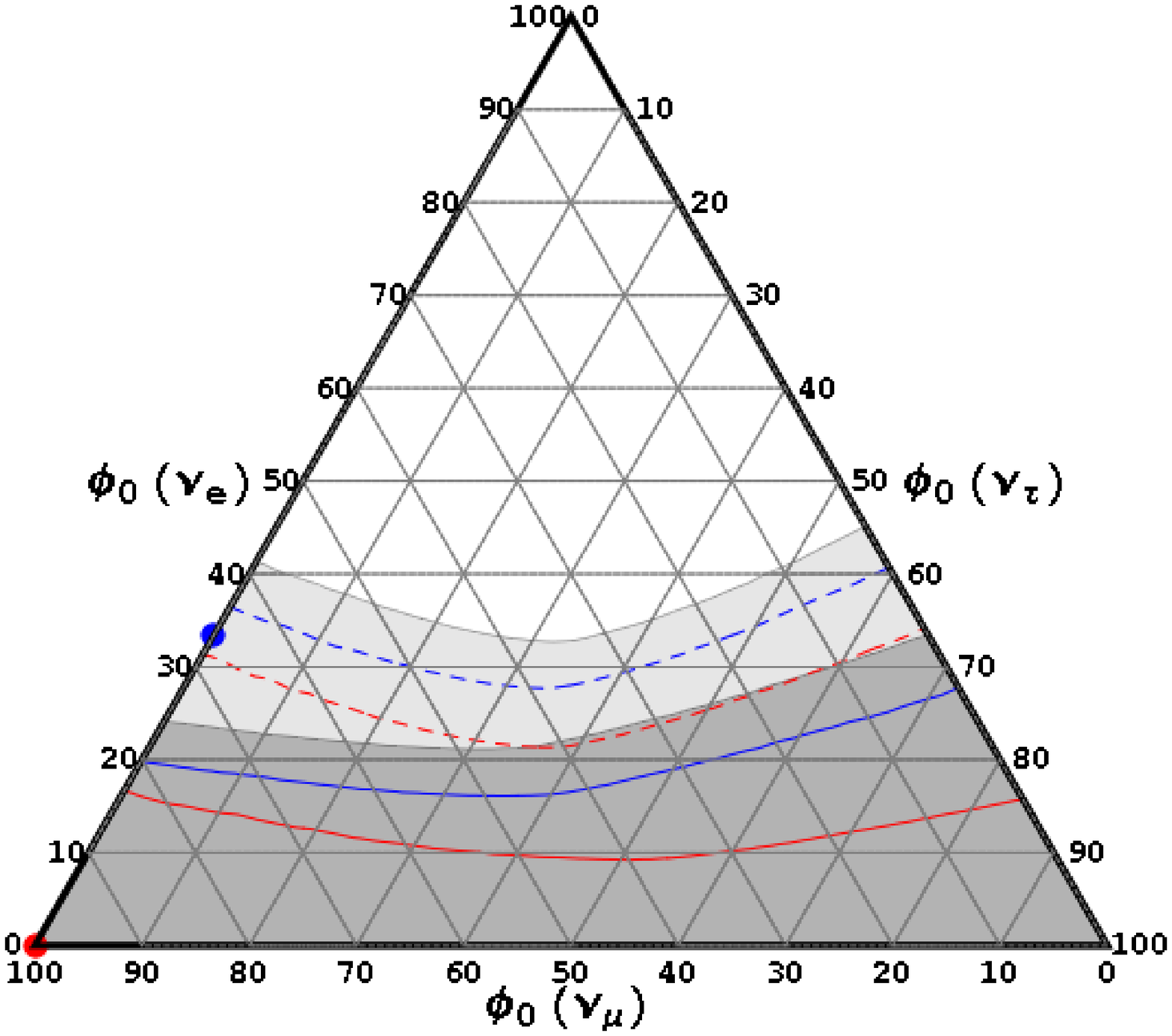}
\includegraphics[scale=0.23]{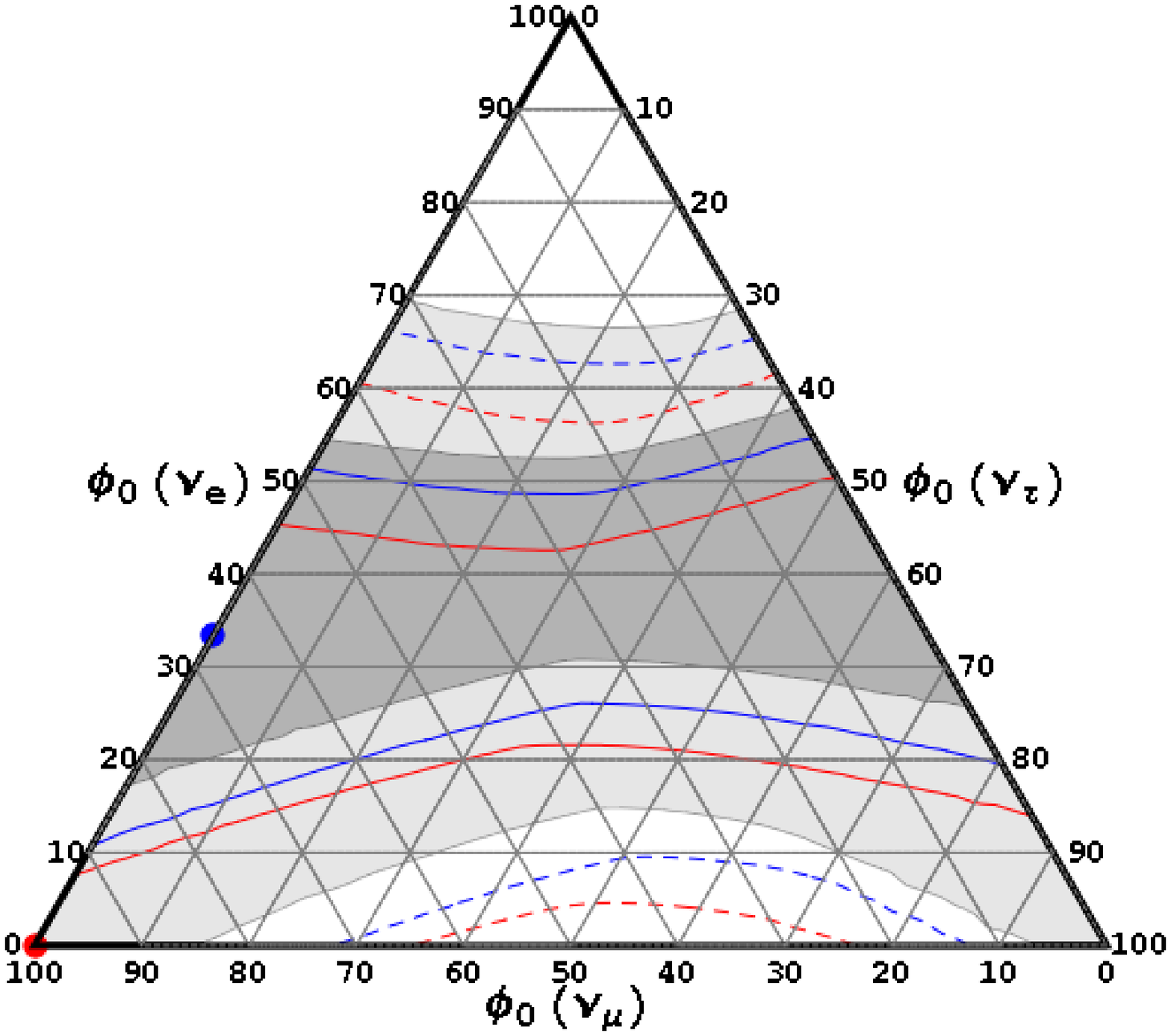}
\includegraphics[scale=0.23]{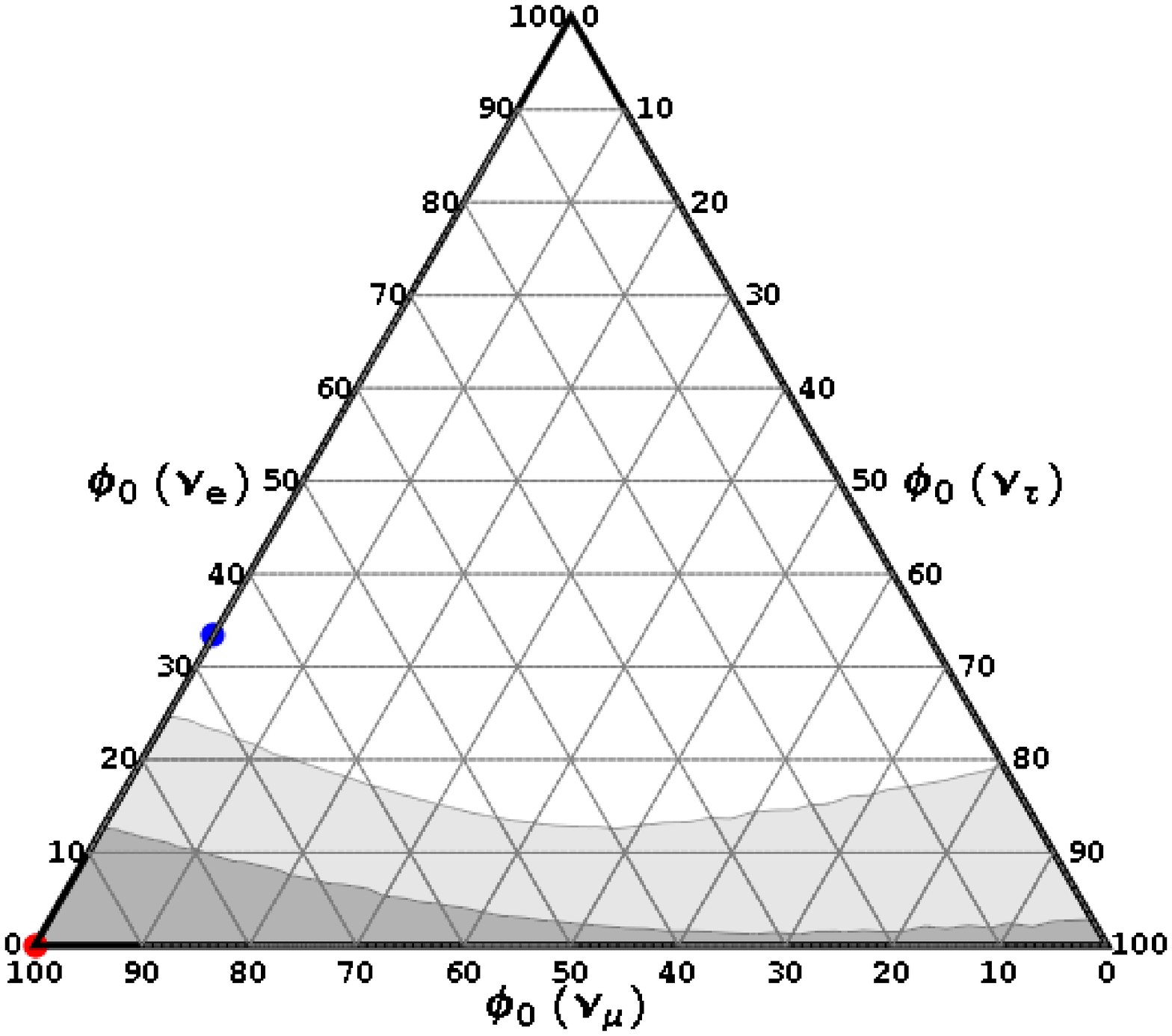}
\includegraphics[scale=0.23]{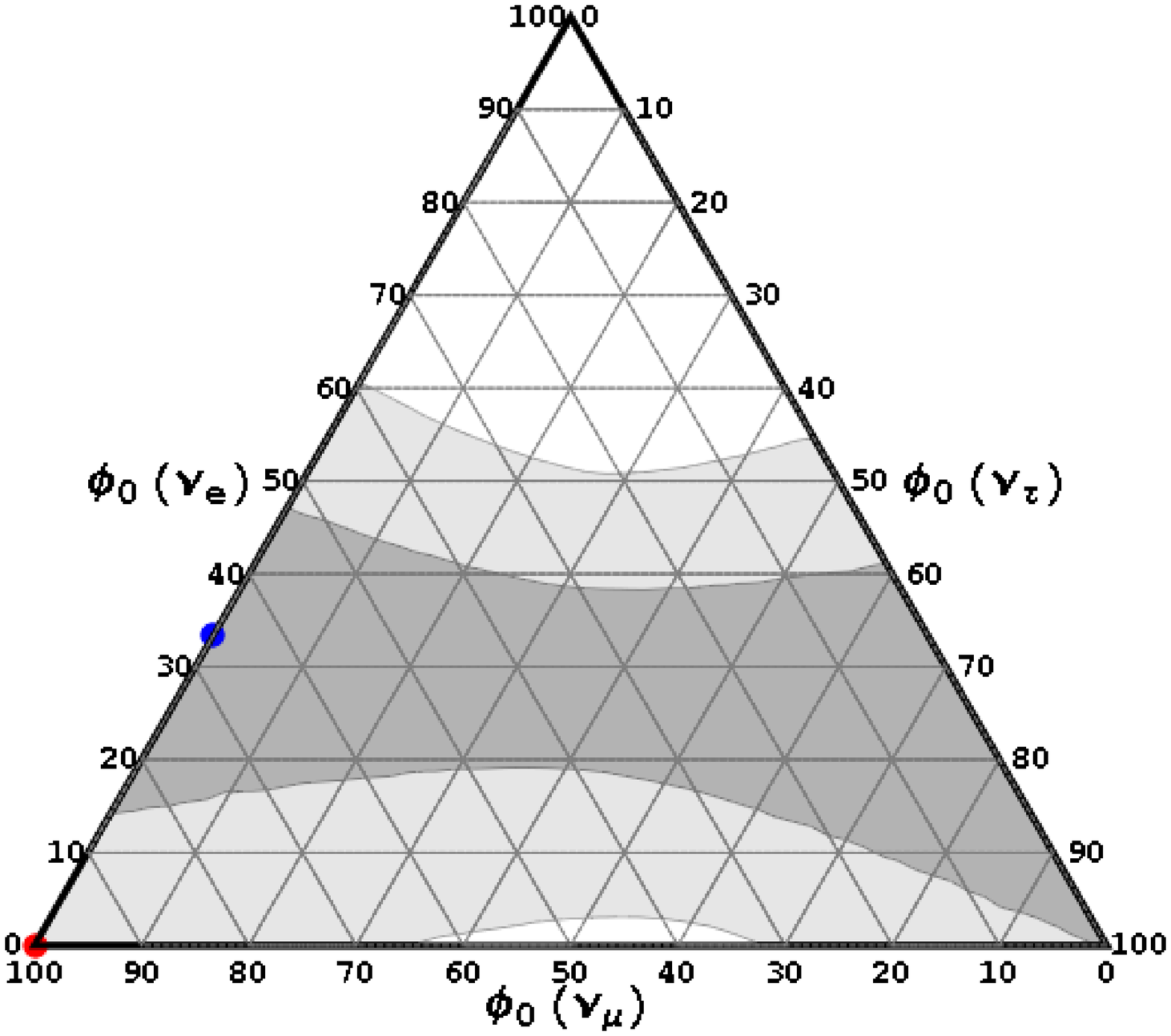}
\caption{The reconstructed range of neutrino flavor ratio with
different values of neutrino mixing angles. The first two panels are
obtained with $\sin^2\theta_{13}=0.016\pm 0.010$ \cite{Fogli}, and
the input $CP$ phase taken to be $0$, $\pi/2$ and $\pi$
respectively. The ranges for $\theta_{12}$ and $\theta_{23}$ follow Eq.~(\ref{bestfit}).
 Light gray area, dashed blue and dashed red lines
correspond to $3\sigma$ ranges for the reconstructed neutrino flavor
ratio at the source for $\cos\delta=1$, $\cos\delta=0$ and
$\cos\delta=-1$ respectively. Gray area, blue and red lines
correspond to $1\sigma$ ranges for the reconstructed neutrino flavor
ratio at the source for $\cos\delta=1$, $\cos\delta=0$ and
$\cos\delta=-1$ respectively. The next two panels are the results
for an input muon-damped source and an input pion source,
respectively, with $\sin^2\theta_{23}=0.55^{+0.09}_{-0.06}$.
 The ranges for $\theta_{12}$ and $\theta_{13}$ follow Eq.~(\ref{bestfit}).
 All of the panels are obtained with $\Delta R/R=10\%$ and $\Delta
S/S=12\%$.} \label{fig2}
\end{figure}

The statistical analysis is performed with the following formula \cite{Blum,Rodejohann}
\be
\chi^2=\left(\frac{R_{\rm th}-R_{\rm exp}}{\sigma_{R_{\rm
exp}}}\right)^2+ \left(\frac{S_{\rm th}-S_{\rm exp}}{\sigma_{S_{\rm
exp}}}\right)^2 +\sum_{jk=12,23,13}
\left(\frac{s_{jk}^2-(s_{jk})^2_{\rm best \,
fit}}{\sigma_{s_{jk}^2}}\right)^2\label{chi}
\ee
with $\sigma_{\rm Rexp}=(\Delta R/R)R_{\rm exp}$, $\sigma_{\rm
Sexp}=(\Delta S/S)S_{exp}$, $s^2_{ij}\equiv \sin^2\theta_{ij}$ and
$\sigma_{s_{jk}^2}$ denote the 1$\sigma$ range of $s^2_{ij}$. The
suffix ``th'' indicates the theoretical predicted value which
depends on the source neutrino flavor ratio and the neutrino mixing
angles in Eq.~(\ref{bestfit}). The suffix ``exp'' indicates the
experimentally measured value which is generated by the true
neutrino flavor ratio and the best-fit values of neutrino mixing
angles. $\Delta R$ and $\Delta S$ are assumed to be dominated by
statistical errors which imply \cite{Blum}
\begin{eqnarray}
\left(\frac{\Delta
S}{S}\right)=\frac{1+S}{\sqrt{S}}\sqrt{\frac{R}{1+R}}\left(\frac{\Delta
R}{R}\right), \label{Poisson}
\end{eqnarray}
It is easier to measure $R$ than to measure $S$. We present results
of neutrino flavor reconstruction in Fig.~\ref{fig1}. The
reconstructed region of neutrino flavor ratio with $\Delta R/R=10\%$
and $\Delta S/S=12\%$ (Poisson relation) is much smaller than the
reconstructed region with the measurement $\Delta R/R=10\%$ only.
For an input muon-damped source, the pion source can be ruled out at
the $3\sigma$ level with both $R$ and $S$ measured to the above
mentioned accuracies. On the other hand, for an input pion source,
the muon-damped source can not be ruled out at the 3$\sigma$ level
under the same condition. The ranges for neutrino mixing angles and
the true value for the $CP$ phase could affect the reconstructed
range for the neutrino flavor ratio. These effects are shown in
Fig.~\ref{fig2}.

\section{Conclusion}
We have illustrated the reconstruction of the neutrino flavor ratio
at the source from the measurements of energy-independent ratios
$R\equiv\phi (\nu_{\mu})/\left(\phi (\nu_{e})+\phi
(\nu_{\tau})\right)$ and $S\equiv\phi (\nu_e)/\phi (\nu_{\tau})$
among integrated neutrino flux. By just measuring $R$ alone from
either an input pion source or an input muon-damped source with a
precision $\Delta R/R=10\%$, the reconstructed $3\sigma$ range for
the initial neutrino flavor ratio covers the entire
physical range for the above ratio. By measuring both $R$ and $S$
from an input muon-damped source, the pion source can be ruled out
at the $3\sigma$ level with $\Delta R/R=10\%$ and $\Delta S/S$
related to the former by the Poisson statistics. The full details of
our studies are presented in \cite{ours}.
\\
\\
\noindent{\bf Acknowledgements}
The authors appreciate supports by National Science Council, Taiwan, under the Grant No. 96-2112-M-009-023-MY3,
Research and Development Office,
National Chiao-Tung University and Leung Center for Cosmology and
Particle Astrophysics, National Taiwan University.


\begin{thebibliography}{99}
\itemsep -2pt
\bibitem{Rachen} J.P. Rachen and P. Meszaros, \Journal{\PRD} {58}{123005} {1998}
\bibitem{Kashti} T. Kashti and E. Waxman, \Journal{\PRL}{59}{181101}{2005}
\bibitem{Kachelriess} M. Kachelriess, S. Ostapchenko and R. Tomas, \Journal{\PRD}
{77} {023007} {2008}
\bibitem{Gonzalez}  M.C Ganzales-Garcia and M. Maltoni, \Journal{\PREP}{460}{1}{2008}
\bibitem{Blum} K. Blum, Y. Nir and E. Waxman, arXiv:0706.2070 [hep-ph]
\bibitem{Rodejohann} S. Choubey, V. Niro and W. Rodejohann,
\Journal{\PRD} {77} {11306}{2008}
\bibitem{Fogli} G.L. Fogli, E. Lisi, A. Marrone, A. Palazzo and A.M. Rotunno, \Journal{\PRL}{101}{141801}{2008}
\bibitem{ours} K.~C.~Lai, G.~L.~Lin and T.~C.~Liu, \Journal{\PRD}{80}{103005}{2009}

\end{thebibliography}
\end{document}